\documentclass[final, twocolumn,5p]{elsarticle}
\usepackage{graphics}
\usepackage{amsmath}
\usepackage{balance}
\usepackage{amssymb}
\usepackage{caption}
\usepackage{multirow}
\usepackage{booktabs}
\usepackage{color}
\usepackage[bottom]{footmisc}
\usepackage{epstopdf}

\usepackage{threeparttable}
\usepackage{tabularx}
\usepackage{booktabs}

\biboptions{square,sort&compress}

\journal{Scripta Materialia}

\begin{document}

\author[turku,kth]{Song Lu\corref{song}}
\cortext[song]{Corresponding author. Tel: +46 8 7906215; Fax:+46 8 100-411}
\ead{lusommmg@hotmail.com}
\author[kth]{Wei Li}
\author[pohang]{Se Kyun Kwon}
\author[turku]{Kalevi Kokko}
\author[imr]{Qing-Miao Hu}
\author[outokumpup]{Staffan Hertzman}
\author[kth,uppsala,hungary]{Levente Vitos}

\address[turku]{Department of Physics and Astronomy, University of Turku, FI-20014 Turku, Finland}%
\address[kth]{Applied Materials Physics, Department of Materials Science and Engineering, Royal Institute of Technology, Stockholm SE-100 44, Sweden}
\address[pohang]{Graduate Institute of Ferrous Technology, Pohang University of Science and Technology, Pohang 37673, Korea}
\address[imr]{Shenyang National Laboratory for Materials Science, Institute of Metal Research, Chinese Academy of Sciences, 72 Wenhua Road, Shenyang 110016, China}%
\address[outokumpup]{Outokumpu Stainless Research Foundation, Avesta Research Center, SE-774 22 Avesta, Sweden}%
\address[uppsala]{Department of Physics and Astronomy, Division of Materials Theory, Uppsala University, Box 516, SE-751210, Uppsala, Sweden}%
\address[hungary]{Research Institute for Solid State Physics and Optics, Wigner Research Centre for Physics, Budapest H-1525, P.O. Box 49, Hungary}%

\date{9 October 2015}

\title{Effect of interstitial-driven lattice expansion on the stacking fault energy in austenitic steels}%

\begin{abstract}
Interstitials (carbon and nitrogen) are crucial alloying elements for optimizing the mechanical performance of the twinning-induced plasticity (TWIP) steels in terms of the stacking fault energy (SFE). First-principles calculations have been performed to study the effect of interstitial-induced lattice expansion on the SFE. Comparing the predictions with the SFEs measured for alloys containing C and N, our results suggest that the dominant effect of these interstitials on the SFE is due to the lattice expansion effect.
\end{abstract}


\maketitle

The stacking fault energy (SFE, $\gamma$) of austenite is an important physical parameter for the transformation-induced plasticity (TRIP) and twinning-induced plasticity (TWIP) steels because it has a great influence on the plastic deformation behavior. Extensive studies have been performed to study the effects of composition, temperature, grain size, strain rate, etc. on the SFE (see Ref.\cite{Saeed2009} and references therein).

Carbon and Nitrogen are the key alloying elements in optimizing the mechanical properties of TRIP/TWIP steels. Summarizing the experimental results for the dependence of the SFE on the C content in steels reveals strikingly inconsistencies (see Fig.1 in Ref.\cite{Hickel2014147}). Similarly, controversial effects of N on the SFE in austenitic steels are reported in literature. The addition of N was reported to  promote the martensitic ($\epsilon$) transformation in Fe-16Mn-(0.015 and 0.05)N (wt.\%) steels which indicates that N addition decreases the SFE \cite{Lee201423}. In FeCrNi alloys, N was often found to decrease the SFE (Ref.\cite{Gavriljuk2006537} and references therein). In particular, the comprehensive study performed by Schramm and Reed \cite{schramm1975} indicated that N decreases the SFE by approximately 77 mJm$^{-2}$ per wt.\%.

In contrast to the above results, the addition of 0.05 to 0.23 wt.\% of N to the austenitic Fe-13Cr-19Mn steel led to an increase of SFE from 32 to 40 mJm$^{-2}$, according to an transmission electron microscopy (TEM) study by Petrov~\cite{Petrovbook}. The SFE values in Fe$-$18Cr$-$10Mn$-$0.2Si$-$0.03C$-$(0.39$-$0.69)N (wt.\%) steels were measured by Lee \emph{et al.} and it was shown to increase from 10.4 to 22.8 mJm$^{-2}$  with increasing N concentration \cite{Lee20103173}. Jiang \emph{et al.} \cite{Jiang19961437} reported that the stacking fault probability decreases with increasing N concentration from 0 to 0.047 wt.\% in  Fe$-$30Mn$-$6Si$-$xN (wt.\%) steels, which suggests that N addition increases the SFE.

Non-monotonic alloying effects were also reported. For example, Fujikura \emph{et al.} found that the probability of the stacking faults in Fe-18Cr-10Ni-8Mn-xN alloys decreased at small N additions (0.2-0.3 wt.\%) and increased at higher N contents \cite{Fujikura1975}. Gavriljuk \emph{et al.} ~\cite{Gavriljuk2006537} investigated the effect of N on the SFE in two different austenitic stainless steels. The SFE in Fe$-$15Cr$-$17Mn$-$xN steels decreased from 26 to 20 mJm$^{-2}$ with increasing N concentration from 0.23 to 0.48 wt.\%, then increased to 40 mJm$^{-2}$ when the N concentration was 0.8 wt\%. However, in Fe-18Cr-16Ni-10Mn-xN steels, SFE increased from 43 to 65 mJm$^{-2}$ with increasing N concentration from 0.08 to 0.4 wt.\%, then decreased to 53 mJm$^{-2}$ when the N concentration was 0.54 wt.\%.

Theoretically, the composition dependence of the SFE may be investigated by thermodynamic or first-principles calculations. Within the thermodynamic approach, the SFE is calculated based the Olson-Cohen model~\cite{Olson19761897}, which separates the formation energy into contributions from the Gibbs energy difference between the hexagonal close-packed (hcp) and face-centered cubic (fcc) phases and the interfacial energy ($\sigma_{\gamma/\epsilon}$) between the fcc and hcp phases.  The interfacial energy, however, has not been well defined and has large uncertainty. Actually, the interfacial energy is often adjusted to reach better agreement with the experimental values \cite{Saeed2009}. Normally, the interfacial energy is accepted in the range of 5$-$27 mJm$^{-2}$. In particular, Pierce \emph{et al.} reported that the interfacial energy ranges from 8 to 12 mJm$^{-2}$  in the TRIP/TWIP steels and from 15 to 33 mJm$^2$  in the binary Fe-Mn alloys \cite{Pierce2014238}.

The composition dependence of the SFE in steels has been addressed in several previous studies using first-principles methods \cite{Vitos2006, vitos2006a, Lu2011}. Most of these theoretical studies focused on substitutional alloying elements (Ni, Cr, Co, Nb, Mn, etc.) but the effect of intersitials on the SFE has not yet been well established. The effect of C on the SFE has been studied in a few recent works \cite{Abbasi20113041, Gholizadeh2013341, Medvedeva2014475}. The effect of N in the Fe-X (Fe-12Mn-xN) system was studied by Kibey \emph{et al.} \cite{Kibey2006fen} using \emph{ab initio} methods and they found that the SFE value increased linearly from -404 to 179 mJm$^{-2}$ with increasing N concentration from 0 to 8 at.\%. This prediction is however far too large when compared to the observations.

Most recently, the effect of N on the SFE was quantitatively evaluated in Fe-15Mn-2Cr-0.6C-xN (wt.\%) TWIP steels using X-ray diffraction (XRD), neutron diffraction (ND), and TEM measurements \cite{Lee201423}. It was shown that the SFE linearly increases up to 0.21\% N and $\delta \gamma/\delta c_N$ was measured to be about 100 mJm$^{-2}$ per wt.\% N. Motivated by this study, here we present a series of first-principles calculations which provide a consistent basis to estimate the concentration dependence of the SFE on C and N interstitials in paramagnetic stainless steels. Previous  studies have shown that a proper description of the paramagnetic structure is crucial for calculating the SFE. ~\cite{Vitos2006} The addition of interstitial elements to paramagnetic Fe alloys, however, may induce complex magnetic changes \cite{Boukhvalov2007, Ponomareva2014}. Here we simplify the problem and focus merely on the lattice expansion effect due to the addition of interstitials. Actually, the thermal lattice expansion has been recognized as the main parameter influencing the temperature dependence of SFE \cite{Ruban2007temperature, Hojjatphd, Guvenc2015}. In particular, the SFE calculated at the experimental volume was shown to have better agreement with experimental data \cite{Ruban2007temperature, Hojjatphd, Kibey2007a}. Since C and N have a strong effect on the volume of steels, it is natural to consider the lattice expansion induced by interstitials as one of the plausible mechanisms responsible for the observed variations of the SFE.

The dependence of the SFE on the concentration of interstitials can be expressed as
\begin{equation}
\gamma(c_{\rm C, N})=\gamma_0+A\times c_{\rm C,N},
\end{equation}
where $\gamma(c_{\rm C, N})$ and $\gamma_0$ are the SFEs with and without interstitials, respectively. $A\equiv\delta{\gamma}/\delta c_{\rm C, N}$ is the coefficient of the linear relationship. $c_{\rm C, N}$ is the concentration of C and/or N in weight percentage. In the present work, the SFEs for three alloys, Fe-22Mn, Fe-15Mn-2Cr and Fe-20Cr-8Ni, are calculated at room temperature with respect to the lattice parameter $a$. The calculated SFE values are fitted by a linear function, \emph{viz.}
\begin{equation}
\gamma(a)=\gamma_{0}^{\prime}+\alpha \times a.
\end{equation}
The composition dependence of SFE is then established through the measured linear lattice expansion coefficient ($\beta$=$\delta$a/$\delta c_{\rm C,N}$) which is defined as
\begin{equation}
a(c_{\rm C, N})=a_0+\beta\times c_{\rm C,N},
\end{equation}
where $a(c_{\rm C,N})$ and $a_0$ are the lattice constants with and  without interstitials, respectively. Combining the above expressions (2) and (3), we can simply estimate the effect of lattice expansion by the interstitials on the SFE,
\begin{equation}
A\approx\alpha \times \beta.
\label{eqa}
\end{equation}

The SFE is calculated using the so-called axial interaction model (AIM)~\cite{Denteneer1987}. In the AIM, we take into account interactions between layers up to the third nearest neighbors (3rd order AIM) which for the intrinsic staking fault energy yields
\begin{equation}
\label{eqAIM}
\gamma^{(3)}=\left(F_{\rm hcp}+2F_{\rm dhcp}-3F_{\rm fcc}\right)/S,
\end{equation}
where $F_{\rm hcp}$, $F_{\rm dhcp}$ and $F_{fcc}$ are the free energies for hcp, double hexagonal close-packed (dhcp) and fcc structures, respectively.  $S$ is area of the stacking fault. The contribution of electronic entropy and lattice vibrational entropy at room temperature to the free energy were verified to be relatively insignificant~\cite{Vitos2006, vitos2006a} and thus are neglected in the present work. Then, only the mean-field magnetic entropy contribution (-T$\Delta$S$^{mag}$) is included.
Both the hcp and dhcp lattices were assumed to have the ideal axial ratio ($c/a$=1.633) and the same volume per atom as that of the fcc lattice. The paramagnetic state was described by the Disorder Local Magnetic Moment (DLM) approximation \cite{Gyorffy1985}. The total energies were calculated using the exact muffin-tin orbitals (EMTO) method ~\cite{Andersen1994, Andersen1998, Vitos2007, Vitos2001b, Vitos2000} in combination with the coherent potential approximation (CPA)~\cite{Soven1967, Vitos2001}. The one-electron Kohn-Sham equations were solved within the scalar-relativistic approximation and the soft-core scheme. The self-consistent calculations were performed within the generalized gradient approximation proposed by Perdew, Burke and Ernzerhof (PBE)~\cite{Perdew1996}. For more calculational details, readers are referred to Refs. \cite{Hojjatphd, Huang201544, wei2015}.

\begin{figure}[tbh!]
	\centering
	\includegraphics[scale=0.3]{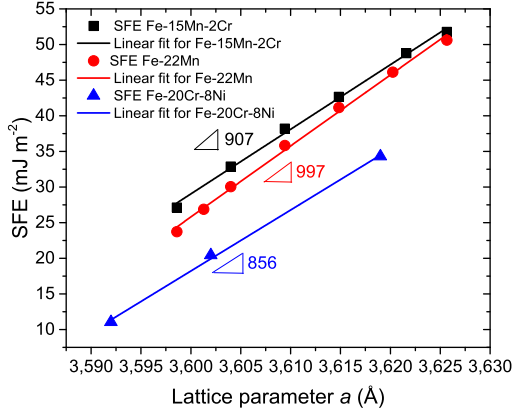}
	\caption{(Color online) The calculated SFEs for paramagnetic Fe-22Mn, Fe-15Mn-2Cr and Fe-20Cr-8Ni alloys with respect to the lattice constant. Results are shown for the room temperature.}
	\label{sfea}
\end{figure}

In Fig.~\ref{sfea}, we plot the calculated SFE at room temperature for Fe-22Mn, Fe-15Mn-2Cr and Fe-20Cr-8Ni alloys as function of the lattice constant. Explicitly, the calculated SFEs for Fe-22Mn and Fe-15Mn-2Cr are larger than the experimental values \cite{Lee201423} (by $\sim$10 mJm$^{-2}$) and previous theoretical results at room temperature \cite{Ruban2007temperature}. This difference may be ascribed to the neglected thermal spin fluctuation \cite{Ruban2007temperature, Hojjatphd} and lattice relaxation \cite{Lee2014588, Pierce2014}. For Fe-20Cr-8Ni, the calculated SFE at room temperature agrees well with the previous theoretical and experimental data \cite{Vitos2006, Lu2011, Hojjatphd}. For all three alloys, the SFE increases linearly with the lattice constant within the present volume interval which includes the measured lattice parameters of these steels. The calculated dependence of SFE on the lattice parameter ($\alpha$) from linear fitting are 997, 907, and 856 mJm$^{-2}$ per \AA~for Fe-22Mn, Fe-15Mn-2Cr and Fe-20Cr-8Ni, respectively.

To transfer the dependence of the SFE on the lattice constant to the dependence on the concentration of interstitials, we collect the available linear lattice expansion coefficients for C and N containing austenites. In general, $\beta$ depends on the steel composition \cite{Scott2007489}. Recently, Lee \emph{et al.} measured the lattice parameters of Fe-15Mn-2Cr-0.6C-xN and Fe-15Mn-xC alloys and for the lattice expansion coefficient $\beta$ for N and C they obtained 0.064 and 0.041 \AA~per wt.\%, respectively \cite{Lee201423, Lee2014588}. These two values are therefore assigned for our Fe-15Mn-2Cr alloy. A linear relation between the lattice parameter of AISI 316 austenitic stainless steel (Fe16-18Cr10-14Ni2-3Mo2Mn) and C concentration at unstrained condition was reported by Hummelsh{\o}j \emph{et al.}~\cite{Hummelshoj2010761}. The corresponding lattice expansion coefficient $\beta$ is 0.028 \AA~per wt.\%. The lattice parameters for Fe-18Cr-15Ni-10Mn-xN alloys were measured by Gavriljuk \emph{et al.} \cite{ Gavriljuk} and the resulted $\beta$ for N is about 0.025 \AA~per wt.\% which is very close to the value derived in Fe-N steels~\cite{Cheng1990509}. These two values are adopted for the present Fe-20Cr-8Ni alloy. For Fe-22Mn, surprisingly, there is no direct lattice expansion coefficients reported for alloys with similar compositions. We therefore take the values derived in $\gamma$-Fe, $\beta\approx$0.029 \AA~per wt.\% for N and $\beta\approx$0.032 \AA~per wt.\% for C~\cite{Cheng1990509}.

\begin{table}[tb!]
\centering
\caption{Theoretical and experimental linear coefficients for the SFE versus lattice constant ($\alpha=\delta\gamma/\delta a$), lattice constant versus concentration ($\beta=\delta a/\delta c_{\rm C,N}$), and SFE versus N or C concentration ($A=\delta\gamma/\delta c_{\rm C,N}=\alpha\times\beta$) relations. The units for $\alpha, \beta$ and $A$ are mJm$^{-2}$ per \AA, \AA~per wt.\%, and mJm$^{-2}$ per wt.\%, respectively.}
\label{table}
\smallskip
\small
\begin{threeparttable}
\begin{tabular}{lcccp{1cm}p{1cm}}
\hline\hline
Alloys                &   $\alpha$ & \multicolumn{2}{c}{$\beta$} & \multicolumn{2}{c}{$A$} \\
\cline{3-4}\cline{5-6}
                         &         & N        & C        & N         & C        \\
\midrule
Fe-20Cr-8Ni     & 856  &   0.025\tnote{g}        &  0.028\tnote{f} & 21, -77\tnote{e}   & 24, 26\tnote{d}, 410\tnote{e}  \\
Fe-22Mn               & 997  & 0.029\tnote{i}         & 0.032\tnote{i}         &  29         &   32, 40\tnote{j}       \\
Fe-15Mn-2Cr    & 907  & 0.064\tnote{a} & 0.041\tnote{c}  & 58, 53-102\tnote{a}, 41\tnote{b}, 44\tnote{h}   & 37, 40\tnote{c}\\
\hline\hline
\end{tabular}
\begin{tablenotes}
\item[a]Ref.~\cite{Lee201423}.      \item[b] In Fe-18Cr-10Mn-0.2Si-0.03C-xN (0.39$\leq$x$\leq$0.69 wt.\%) ~\cite{Lee20103173}.
\item[c]Combined effect of C+N on the SFE of Fe-18Cr-10Mn~\cite{Lee2012}.      \item[d] Ref.~\cite{Brofman1978}.
\item[e]Ref.~\cite{schramm1975}. \item[f]Ref.~\cite{Hummelshoj2010761}.   \item[g]Ref.~\cite{JACK19731, Gavriljuk}. \item[h] In Fe-13Cr-19Mn~\cite{Petrovbook}.
\item[i]Ref.~\cite{Cheng1990509} \item[j]Thermodynamic result \cite{Saeed2009}.
\end{tablenotes}
\end{threeparttable}
\end{table}

With the calculated $\alpha=\delta{\gamma}/\delta a$ and experimental lattice expansion coefficient $\beta=\delta a/\delta c_{\rm C,N}$, we can simply estimate the lattice expansion effect induced by the additions of interstitials on the SFE according to Eq. (\ref{eqa}). Our results and the corresponding experimental lattice expansion coefficients are summarized in Table~\ref{table}.

In Fe-20Cr-8Ni, SFE is estimated to increase by $\sim$21 mJm$^{-2}$ per wt.\% for N and by $\sim$24 mJm$^{-2}$ per wt.\% for C.
For comparison, C is obtained to increase SFE by 26 mJm$^{-2}$ per wt.\% in Fe-Cr-Ni alloys according to a four-dimensional linear regression analysis by Brofman and Ansell \cite{Brofman1978}. However, significantly different values were also reported by Schramm and Reed, \emph{i.e.} $A$=-77 mJm$^{-2}$ per wt.\%  for N and $A$=410 mJm$^{-2}$ per wt.\% for C~\cite{schramm1975}, which however was strongly criticized by Brofman and Ansell for large extrapolation errors~\cite{Brofman1978}.

In Fe-22Mn, $A$ for N and C are calculated to be 29 and 32 mJm$^{-2}$ per wt.\%, respectively. The calculated $A$ for C is in nice agreement with previous thermodynamic model result ($A$=40  mJm$^{-2}$ per wt.\%)~\cite{Saeed2009}.
In Fe-15Mn-2Cr, both N and C additions lead to an increase of the SFE, which in general agrees with the observed effect of N and C on stabilizing the austenitic phase and changing the plastic mechanism from martensitic transformation to twinning in TRIP/TWIP steels.

\begin{figure}[tbh!]
	\centering
	\includegraphics[scale=0.3]{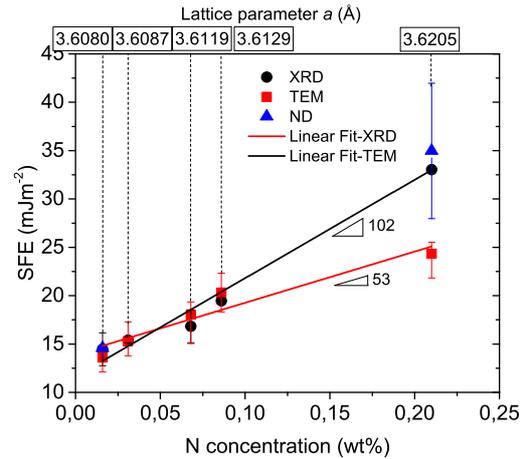}
	\caption{(Color online) The measured SFE for Fe-15Mn-2Cr-0.6C-xN (wt.\%) alloys as a function of N concentration $x$. Data reproduced from Ref.\cite{Lee201423}.}
	\label{expsfe}
\end{figure}

In order to give a direct comparison, the measured SFE of Fe-15Mn-2Cr-0.6C-xN with respect to N content from Ref.\cite{Lee201423} is re-plotted in Fig.~\ref{expsfe}. Both XRD and TEM results indicate a linear increase of SFE with respect to N concentration. XRD and TEM give very similar results for N content up to 0.1 wt.\%. For 0.21 wt.\% N, TEM result is much smaller than the XRD and ND results.  Linear fit for the XRD and TEM SFE data results two different composition dependence ratios,  $A$=102 and 53  mJm$^{-2}$ per wt.\%, respectively. It was argued by the authors that the TEM is not suitable for measuring SFE higher than approximately 20 mJm$^{-2}$ due to the difficulty in accurate measurement of the inscribed radius of the extended dislocation node. Discarding the results for 0.21 wt.\% N, $A$ is about 78 mJm$^{-2}$ per wt.\%. Our predicted $A$ for N is in a nice agreement with the experimental value determined by TEM, but smaller than the values by XRD~\cite{Lee201423}.

The SFE depends strongly on the local concentrations of interstitials near stacking faults, other than the overall concentration~\cite{Hickel2014147}. However, it is still controversial  regarding the segregation or partitioning of interstitials to the stacking fault in literature.  Petrov \emph{et al.} observed significant C segregation to internal surfaces  in Fe-22Mn-0.69C and the measured SFE first decreases and then increases with increasing C concentration~\cite{Petrov19931471}.  Seol \emph{et al.} found that in the Fe-Mn steels C can be trapped at phase boundaries between austenite and $\epsilon$-martensite, stacking faults in austenite and prior austenite grain boundaries~\cite{Seol2013248}. However, a very recent study on the effect of nanodiffusion of C on SFE showed that the TEM method for measuring SFE may locally heat the sample and trigger the outward diffusion of C atoms from stacking fault to fcc matrix \cite{Hickel2014147, Herbig201537}.
Previous \emph{ab initio} calculations have shown that the chemical effect of C is to strongly increase SFE only when C atoms are very close to stacking faults (within 1-2 atomic layer distance).  Depletion of C near stacking fault therefore decreases the measured SFE by TEM~\cite{Hickel2014147}. Similar mechanism can also be applied for N, which may explain the smaller SFE by TEM than the XRD and ND results in the Fe-15Mn-2Cr-xN alloys (Fig.\ref{expsfe}). The depletion of N at stacking fault results in the dissipation of chemical effect and leaves mainly the lattice expansion effect. In this respect, the fact that our predicted $A$  based merely on volume effect is in good agreement with the experimental $A$ from fitting TEM SFE values in Fe-15Mn-2Cr-0.6C-xN gives an indirect evidence supporting the above nanodiffusion mechanism.

In other Fe-Cr-Mn steels, the measured $A$ values are of similar magnitude. Lee \emph{et al.} reported that the SFE values measured by neutron diffraction  increase linearly from 10.4 to 22.8 mJm$^{-2}$ with increasing N concentration from 0.39 to 0.69 wt.\% in Fe-18Cr-10Mn-0.2Si-0.03C-xN (wt.\%) steels~\cite{Lee20103173}. The corresponding $A$ is about 41 mJm$^{-2}$ per wt.\% N. Unfortunately, they didn't report the change of lattice parameter with respect to N content. Petrov \emph{et al.} reported a similar value in Fe-13Cr-19Mn steel ($A\sim$44 mJm$^{-2}$ per wt.\% N)~\cite{Petrovbook}. The $A$ for C in Fe-15Mn-2Cr is calculated to be 37 mJm$^{-2}$ per wt.\% . The combination effect of C and N on SFE in Fe-18Cr-10Mn was studied by means of ND and TEM~\cite{Lee2012}. Three-dimensional linear regression analysis showed that SFE is increasing with C+N concentration by $\sim$ 40  mJm$^{-2}$ per wt.\% (C+N).

In summary, the above results demonstrate that a very large portion of the effect of interstitials on SFE is due to their effects on the lattice parameter. The deviation of our results from the experimental values may be attributed to several aspects, such as the chemical and magnetic effects of interstitials on the SFE, the differences in the chemical compositions of the steels used, the range of N concentrations studied, the measuring equipment used as well as the theoretical and experimental error bars.

Valuable discussions with Erik Schedin and Jan Y. Jonsson at Outokumpu Avesta Research Center are highly appreciated. This work was supported by the Swedish Research Council, the Swedish Foundation for Strategic Research, the Carl Tryggers Foundation, the Chinese Scholarship Council and the Hungarian Scientific Research Fund (OTKA 84078 and 109570). Qing-Miao Hu acknowledges the financial support from the NSF of China under grant no. 51271181 and no. 51171187 and the Chinese MoST under grant no. 2014CB644001. Song Lu acknowledges Magnus Ehrnrooth foundation for providing a Postdoc. Grant. We acknowledge the Swedish National Supercomputer Centre in Link\"oping for computer resources.

\balance

\end{document}